\documentclass[referee,pdflatex,sn-mathphys]{sn-jnl}

\usepackage{graphicx}%
\usepackage{multirow}%
\usepackage{amsmath,amssymb,amsfonts}%
\usepackage{amsthm}%
\usepackage{mathrsfs}%
\usepackage[title]{appendix}%
\usepackage{xcolor}%
\usepackage{textcomp}%
\usepackage{manyfoot}%
\usepackage{booktabs}%
\usepackage{algorithm}%
\usepackage{algorithmicx}%
\usepackage{algpseudocode}%
\usepackage{listings}%
\DeclareMathOperator{\R}{{\mathbb R}}
\DeclareMathOperator{\N}{{\mathbb N}}
\DeclareMathOperator*{\argmax}{arg\,max}

\DeclareMathOperator{\E}{{\mathbb E}}
\DeclareMathOperator{\calS}{{\mathcal{S}}}
\DeclareMathOperator{\calX}{{\mathcal{X}}}

\begin{document}

\title[Regionalization approaches for the spatial analysis of extremal dependence]{Regionalization approaches for the spatial analysis of extremal dependence}

\author*[1,2]{\fnm{Justus} \sur{Contzen}}\email{jcontzen@awi.de} % orcid: 0000-0003-0936-9376

\author[3]{\fnm{Thorsten} \sur{Dickhaus}} % orcid: 0000-0003-3084-3036

\author[1,2]{\fnm{Gerrit} \sur{Lohmann}} % orcid: 0000-0003-2089-733X

\affil[1]{\orgdiv{Section Paleoclimate Dynamics}, \orgname{Alfred Wegener Institute Helmholtz Center for Polar and Marine Research}, \orgaddress{\city{Bremerhaven}, \country{Germany}}}

\affil[2]{\orgdiv{Department of Environmental Physics}, \orgname{University of Bremen}, \orgaddress{\city{Bremen}, \country{Germany}}}

\affil[3]{\orgdiv{Institute for Statistics}, \orgname{University of Bremen}, \orgaddress{\city{Bremen}, \country{Germany}}}

\abstract{The impact of an extreme climate event depends strongly on its geographical scale. Max-stable processes can be used for the statistical investigation of climate extremes and their spatial dependencies on a continuous area. Most existing parametric models of max-stable processes assume spatial stationarity and are therefore not suitable for the application to data that cover a large and heterogeneous area. For this reason, it has recently been proposed to use a clustering algorithm to divide the area of investigation into smaller regions and to fit parametric max-stable processes to the data within those regions. We investigate this clustering algorithm further and point out that there are cases in which it results in regions on which spatial stationarity is not a reasonable assumption. We propose an alternative clustering algorithm and demonstrate in a simulation study that it can lead to improved results.}


%%\pacs[MSC Classification]{62H12, 62H30}

\maketitle

\section{Introduction}\label{sec:intro}

Extreme climate events like droughts, heatwaves or heavy rainfall events usually do not occur in isolation at a certain observed location, but in combination with more or less extreme events in the vicinity. Having information on how strongly extremes at different locations influence each other is of high interest as the consequences of large-scale extreme weather are often especially severe. Resources to mitigate the impacts of such events might be limited, while their effects on the economy, the environment or the healthcare system can be drastic. When analyzing climate data in a certain region with regard to extreme events, it is therefore important to focus not only on univariate distributions (i.e. distributions at one location alone), but also on their interdependence and on the multivariate distribution of the data. \\

One common approach in extreme value analysis is the investigation of the block-wise maxima of given time series (for example, annual maxima of daily data). For the spatial analysis of data in a certain region, stochastic processes on a compact subset of $\R^2$ with almost surely continuous sample paths are often employed, which makes it possible to investigate spatial dependencies not only for the locations of the spatial data at hand, but also for unobserved locations in the area. Under mild regularity conditions, spatial block-wise maxima can be described using max-stable processes. Using a spectral representation, parametric sub-classes of max-stable processes can be derived which allow for parametric inference. A composite maximum likelihood estimator can be used to fit such processes to given data. Different sub-classes of max-stable processes have been proposed \citep{schlather, brown-resnick, kabluchko, Opitz_2013}, but they all describe processes that are spatially stationary, i.e. their finite-dimensional marginal distributions are invariant with respect to translations in space. Such an assumption may be reasonable if the investigated area is small and rather homogeneous, and indeed such models have successfully been applied to climate extremes in Switzerland \citep{ribatet}. For the application to data from a larger area of investigation they are not well suited, and non-stationary extensions of them are an active research topic. \cite{husergenton} propose such a max-stable model based on non-stationary Gaussian processes. Their model flexibly captures non-stationarity, but it relies on covariates which are not given for every data set. Another approach to investigate large-scale data using max-stable processes is due to \cite{saunders}. They employ a clustering algorithm by \cite{bernard} to split the investigated area into several small regions. To each of the smaller regions, a stationary max-stable process is then fitted. \\

In this work, we will build up on the aforementioned clustering approach. While clustering is a promising idea to make stationary max-stable processes applicable, we will see that the algorithm by \cite{bernard} does not necessarily produce regions in which spatial stationarity is a reasonable assumption. We will illustrate this using a simple example of a non-stationary max-stable process, and we will propose an alternative clustering algorithm and compare the two in a simulation study. \\

The rest of this paper is structured as follows: In the next section, we will give a brief introduction to max-stable processes. The clustering algorithm by \cite{saunders} and our suggested algorithm are discussed in Section \ref{sec:clustering}. Using a simulation study, we investigate the performance of the two algorithms in Section 
\ref{sec:simulation}. Section \ref{sec:discussion} on conclusions and a discussion finalize the article.

\section{Theoretical foundations}
\label{sec:theory}
This section gives a brief overview of the theory of max-stable processes, for a more detailed treatment see \cite{dehaan_ferreira} and \cite{ribatet}. Let $\calS$ be a compact subset of $\R^2$. Throughout the paper, we will investigate stochastic processes on $\calS$ with almost surely continuous sample paths and we use the short notation $Y$ for $\{Y(s)\}_{s\in \calS}$. The distribution of such a process is completely determined by its finite-dimensional marginal distributions. \\

A stochastic process $Y$ is called max-stable if for all $n \in \N$ there exist continuous functions $\{c_n(s)\}_{s \in \calS}$ and $\{d_n(s)\}_{s \in \calS}$ with $c_n(s) > 0$ such that for independent copies $Y_1,\dots,Y_n$ of $Y$
\begin{align}
    \Big{\{}\frac{\max_{i=1,\dots,n} Y_i(s) - d_n(s)} {c_n(s)}\Big{\}}_{s \in \calS} \overset{\mathcal{D}}{=} \{Y(s)\}_{s \in \calS}
\end{align}
holds, with $\overset{\mathcal{D}}{=}$ denoting equality in distribution. For $n \in \N$, let $X_1,\dots,X_n$ be independent and identically distributed stochastic processes and define $M_{(n)}(s) := \max_{i=1,\dots, n}(X_i(s))$. We assume the existence of functions $a_n(\cdot) > 0$ and $b_n(\cdot)$ on $\calS$ such that $M_{(n)}$ converges in distribution ($\xrightarrow{\mathcal{D}}$) to some process $Y$ as $n$ tends to infinity:
\begin{align}
\label{gev_def}
\Big{\{}\frac {\{M_{(n)}(s) - b_n(s)} {a_n(s)}\Big{\}}_{s \in \calS} \xrightarrow{\mathcal{D}} \{Y(s)\}_{s \in \calS}. 
\end{align}
Then, $Y$ must be a max-stable process if it is not degenerate. In practical applications, the data are grouped into blocks of a fixed block size $n$, and it is assumed that a max-stable process is a reasonable approximation to the resulting block-wise maxima. The margins of a max-stable process are necessarily max-stable random variables, and it is well known that as such, they must follow a generalized extreme value (GEV) distribution. The GEV distributions form a parametric family for which statistical inference is well-established and several methods of parameter estimation exist \citep[see e.g.][Chapter 7]{mcneil}. Using marginal transformations, it can be assumed without loss of generality that all marginal distributions of the investigated max-stable processes are identical (note that the GEV distributions are absolutely continuous), and it is common to consider margins that are unit Fréchet distributed (cdf $\Phi$ given by $\Phi(z) = \exp(-z^{-1})$, $z > 0$). A max-stable process with unit Fréchet margins is called a simple max-stable process. \\

The finite-dimensional marginal distribution of a simple max-stable process $Y$ at locations $s_1,\dots,s_n \in \calS$, $n \in \N$, can be written as
 \begin{align}
     \Pr(Y(s_1) \leq y_1,...,Y(s_n) \leq y_n) = \exp(-V_{s_1,\dots,s_n}(y_1,\dots,y_n))
    \label{eq:v}
 \end{align}
for a function $V_{s_1,\dots,s_n}$ that fulfills $V_{s_1,\dots,s_n}(ay_1,\dots,ay_n) = a^{-1}V_{s_1,\dots,s_n}(y_1,\dots,y_n)$ for all $a > 0$ \citep{dehaan_ferreira}. In practical applications, a max-stable process is fitted to data that are given on a finite set of locations $\calX \subseteq \calS$ (for example the locations of weather stations). The data at each location have been transformed to a unit Fréchet distribution. Denote the data at location $x \in \calX$ by $y_x^{(1)},\dots,y_x^{(m)}$ with $m \in \N$ the sample size. Below, we will present several parametric models for max-stable processes, but before doing so, we will discuss how to fit  parametric models to data. Assume a parametric subclass of max-stable processes with a parameter space $\Psi$, which implies in particular that the functions $V_{s_1,\dots,s_n}$ are parametrized as $V_{s_1,\dots,s_n}^\psi$ with $\psi \in \Psi$ for all $s_1,\dots,s_n \in \calS$. The multivariate density of a max-stable process $Y$ at $s_1,\dots,s_n \in S$ can be derived from Eq.~\eqref{eq:v} by taking the mixed derivative $\partial s_1 \dots \partial s_n$. It contains a sum of products of partial derivatives of $V_{s_1,\dots,s_n}^\psi$. Even for a moderate value of $n$ like $n = 10$, the number of summands in the density is so large that numerical maximization is not computationally feasible. The classical maximum-likelihood approach is therefore usually not applicable to the multivariate density of all data points in $\calX$ \citep{ribatet}. As a remedy, it is common to use a composite likelihood approach instead: Maximum-likelihood estimation is not carried out by maximizing the full log-likelihood, but instead by maximizing the sum of bivariate marginal log-likelihoods 
\begin{align}
\hat\psi = \argmax_{\psi \in \Psi} L(\psi)
\end{align}
with
\begin{align}
L(\psi) = \sum_{i=1}^m \sum_{x_1 \neq x_2 \in \calX} \log f^\psi_{x_1,x_2}(y_{x_1}^{(i)},y_{x_2}^{(i)}).
\label{eq:composite_llh}
\end{align}
The bivariate densities can be calculated as
\begin{multline}
 f^\psi_{x_1,x_2}(y_1,y_2) = \exp\Big(\! -V^\psi_{x_1,x_2}(y_1,y_2)\Big) \, \cdot \\ \qquad \Big(\frac \partial {\partial y_1}V^\psi_{x_1,x_2}(y_1,y_2)\frac \partial {\partial y_2}V^\psi_{x_1,x_2}(y_1,y_2) -
 \frac {\partial^2} {\partial y_1 \partial y_2}V^\psi_{x_1,x_2}(y_1,y_2)\Big).
\end{multline}

To reduce numerical complexity, it is also common practice to include only those pairs $(x_1,x_2)$ in Eq.~\eqref{eq:composite_llh} for which $\|x_1-x_2\|$  does not exceed a certain threshold (throughout the paper, $\|\cdot\|$ denotes the Euclidean norm). \\

As mentioned in the introduction, we will use clustering algorithms to split the set $\calX$ into different subsets and and we will fit max-stable processes to the data on each subset. Introducing the notation for this, let $M \subseteq \calX$ be a subset of $\calX$ with at least two elements, then we write the corresponding composite likelihood as
\begin{align}
    L_M(\psi) = \sum_{i=1}^m \sum_{x_1 \neq x_2 \in M} \log f^\psi_{x_1,x_2}(y_{x_1}^{(i)},y_{x_2}^{(i)})
\label{eq:llh_m}
\end{align}
and the corresponding estimator as
\begin{align}
    \hat\psi_M = \argmax_{\psi \in \Psi} L_M(\psi).
\end{align} \\

Parametric subfamilies of max-stable processes are constructed using the following stochastic representation \citep{dehaan, penrose}: Let $Y$ be a simple max-stable process. Then,
\begin{align}
\label{eq:spectr}
\Big\{Y(s)\Big\}_{s \in \calS} \overset{\mathcal{D}}= \Big\{ \max_{i \geq 1} \zeta_i Z^{(i)}(s) \Big\}_{s \in \calS}, \end{align}
where $\{\zeta_i\}_{i \in \N}$ are the points of a Poisson point process with intensity measure $d\Lambda(\zeta) = \zeta^{-2} d\zeta$ and $\{Z^{(i)}\}_{i \in \N}$ are independent copies of a non-negative stochastic process $Z$ fulfilling $\E[Z(s)] = 1$ for all $s \in \calS$. The process $Z$ is called spectral process of $Y$. \\

Thus, the distribution of $Y$ can be modeled by choosing a model for $Z$. One of the first parametric subfamilies proposed was the Schlather model \citep{schlather}, using the underlying process $Z$ given by
\begin{align} Z(s) = \sqrt{2\pi} \max(0, G(s)) 
\end{align}
with $G$ a standard Gaussian process with a spatially stationary covariance function $\rho(h)$ (depending only on $h = s_1 - s_2 \in \R^2$, $s_1, s_2 \in \calS$). \cite{Opitz_2013} showed that the Schlather process has a limited scope of applicability because its bivariate distributions always exhibit extremal dependence, even if the underlying Gaussian variables are uncorrelated. To avoid this limitation, he extended Schlather's model to the extremal-t model, using the spectral process $Z$ given by
\begin{align}
Z(s) = c_\nu \max(0, G(s))^\nu
\end{align}
with $\nu \geq 1$ fixed, $G$ as in the Schlather model and $c_\nu$ a constant ensuring that $\E[Z(s)]=1$. For this process, it holds
\begin{multline}
    V_{s, s+h}(y_1,y_2) = \frac 1 y_1  T_{\nu+1}\Big(\frac{(y_2/y_1)^{1/\nu}-\rho(h)} {(\nu+1)^{-1/2}(1-\rho(h)^2)^{1/2}}\Big) + \\
     \qquad\qquad \frac 1 y_2  T_{\nu+1}\Big(\frac{(y_1/y_2)^{1/\nu}-\rho(h)} {(\nu+1)^{-1/2}(1-\rho(h)^2)^{1/2}}\Big)
\end{multline}
with $T_{\nu+1}$ the cdf of the Student-t distribution with $(\nu+1)$ degrees of freedom. Bivariate densities can be calculated based on this equation. \\

One possible choice for $\rho$ is $\rho(h) = \exp(-\| h/\lambda \| ^\alpha)$ with $\lambda > 0$ and $0 < \alpha \leq 2$ \citep{ribatet}. To model anisotropic behavior, we follow \cite{davis} and use 
\begin{align}
\rho(h) = \exp(-\|A h\| ^\alpha)  
\end{align}
instead, with $A$ a $2\times 2$ transformation matrix, meaning that
\begin{align}
A = \begin{pmatrix} 
\sin(\gamma)/a  &\cos(\gamma)/(a+b) \\
-\cos(\gamma)/(a+b)  &\sin(\gamma)/b 
\end{pmatrix}
\end{align}
with parameters $a > 0$, $b \geq 0$ and $\gamma \in [0,\pi)$. The level sets of $\rho$ are then ellipses with $a$ and $a+b$ proportional to the lengths of the minor and major axes and $\gamma$ the angle between the horizontal axis and the major axis of the ellipsis. In climate data, elliptical level sets are preferable over the circular ones of isotropic covariance functions because the presence of ocean currents, a predominant wind direction or topographical boundaries frequently cause the extremal dependence structure of two points to depend not only on their distance, but also on the direction.

\section{Clustering algorithms}
\label{sec:clustering}
\subsection{Extremal Dependence Clustering}
As mentioned in the introduction, one approach to investigate data for which the assumption of spatial stationarity is not reasonable is using a clustering algorithm that splits the area of investigation into smaller regions. Stationary max-stable processes like the extremal-t process can then be fitted to the data within those regions. This approach is due to \cite{saunders}, and they perform the regionalization using a hierarchical clustering algorithm (see their paper for an introduction to hierarchical clustering). To apply a hierarchical clustering algorithm, a dissimilarity measure for all pairs of elements in $\calX$ is required. A dissimilarity measure is a non-negative symmetric function $D: \calX \times \calX \rightarrow \R$ that fulfills $D(x,x)=0$ for all $x \in \calX$. \cite{saunders} propose to use a dissimilarity measure first developed by \cite{bernard} that is based on pairwise extremal coefficients, which in turn are defined as
\[ \theta_{x_1,x_2} = V_{x_1,x_2}(1,1). \]
The extremal coefficient is a useful summary measure for the dependency of $Y(x_1)$ and $Y(x_2)$. It takes values between one and two, with a value of one corresponding to the variables being comonotonic and a value of two corresponding to them being stochastically independent, and it can be estimated from a data sample by using the madogram estimator $\hat\theta$ by \cite{ribatet_dombry} and \cite{cooley}.
Defining
\begin{equation}
D_1(x_1,x_2) := \hat\theta_{x_1,x_2} - 1
\end{equation}
yields a dissimilarity function as required. In the following, we will use the term Extremal Dependence Clustering (EDC) for the clustering based on this dissimilarity function. 

\subsection{Applicability of the Extremal Dependence Clustering}
\label{sec:apple}
The EDC algorithm is performed with the goal of defining regions that are suitable for fitting spatially stationary max-stable processes to the data within them. It should therefore group points together in such a way that within the clusters stationarity can be assumed, or is at least a reasonable approximation. The dissimilarity measure by \cite{bernard} is based on the comparison of extremal coefficients and it therefore groups together points with a tendency for concurrent extremes. Within the resulting regions, pairwise extremal dependencies tend to be high in general, which might reduce the possible extent of spatial non-stationarity. Nevertheless, spatial stationarity is not a justified assumption within the clusters defined that way, and the dissimilarity measure by \cite{bernard} was not designed with the intention of finding such regions. \\

We illustrate this using a concrete example of a non-stationary max-stable process. We can construct such a process using an approach by \cite{husergenton}. They extended the stationary extremal-t process by using a non-stationary Gaussian process as underlying spectral process: Instead of \textit{one} $2\times2$ transformation matrix $A$ for the whole space, they use for each point $s \in \calS$ a matrix $A_s$ such that the map $s \mapsto A_s$ is continuous. Let as before $0 < \alpha \leq 2$ be fixed. Using the notations $\Omega_s = (A_s^T A_s)^{-1}$ with $A^T$ denoting the transpose of $A$ and $R_\alpha(x) = \exp(-x^\alpha)$ for $x \in \R_{\geq 0}$, they show how to construct a non-stationary Gaussian process using kernel convolution \citep{paciorek}. The covariance structure of the resulting process is given by
\begin{multline}
    \label{kernel}
\rho(s_1,s_2) = |\Omega_{s_1}|^{\frac 1 4} |\Omega_{s_2}|^{\frac 1 4} \Big|\frac {\Omega_{s_1} + \Omega_{s_2}} 2  \Big|^{-\frac 1 2} \cdot \\
R_\alpha \Big( \sqrt{ ({s_1}-{s_2})^T \Big(\frac {\Omega_{s_1} + \Omega_{s_2}} 2  \Big)^{-1} ({s_1}-{s_2}) } \Big),
\end{multline}
which reduces to the stationary extremal-t process from the previous section if $A_s$ is constant on the whole space. \\

\begin{figure}
    \centering
    \includegraphics[width=\textwidth]{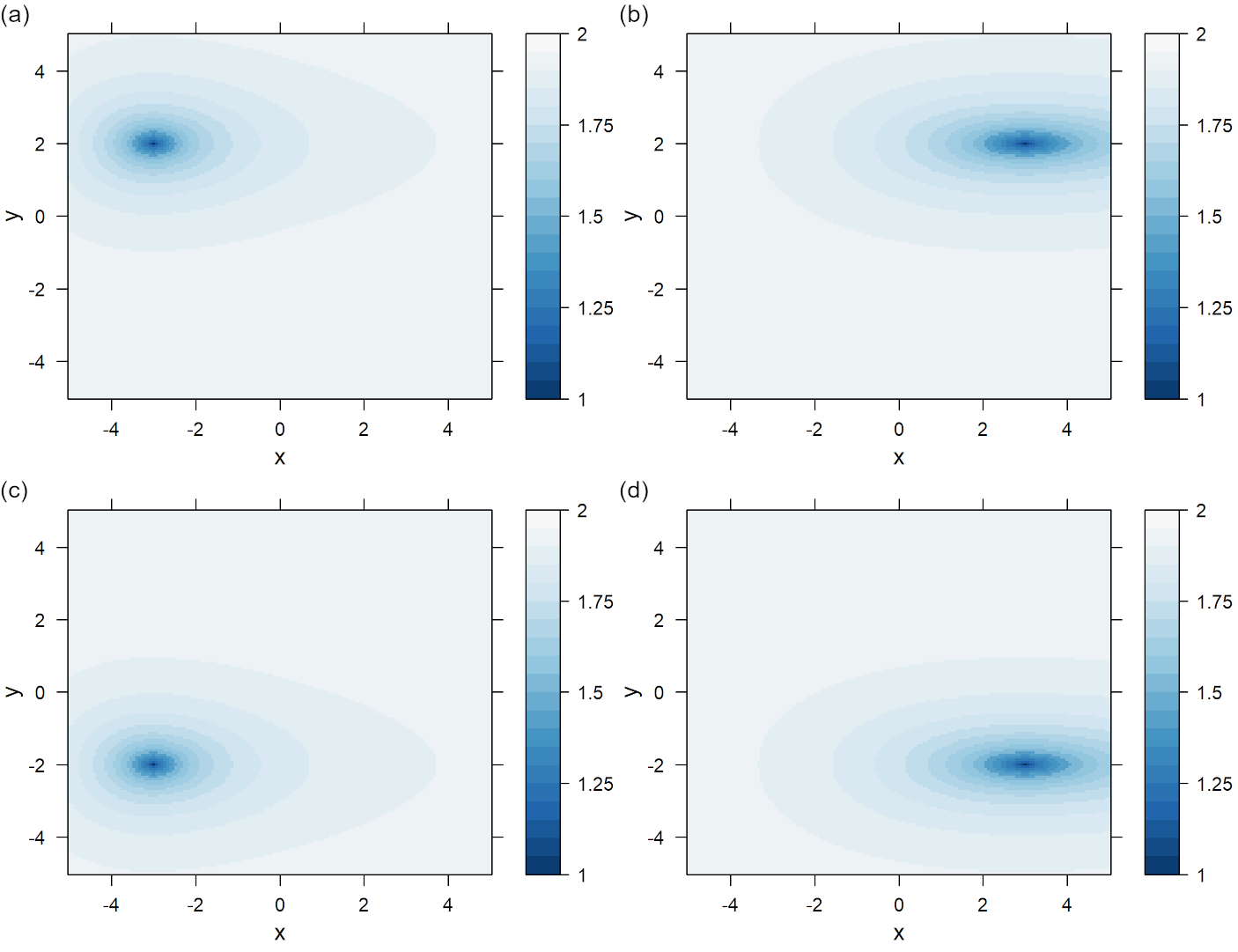}
    \caption{Illustration of the non-stationary max-stable process on $\calS =[-5,5]\times[-5,5]$ given by the matrix parameters $a_s = 2$, $b_s = (x+5)/2$, $s = (x,y) \in \calS$ and $\gamma_s = 0$ and the global parameters $\nu = 5, \alpha = 1$. Displayed are the pairwise extremal coefficients for the points in $\calS$ relative to  \textbf{(a)} $s_1 = (-3, 2)$, \textbf{(b)} $s_2 = ( 3, 2)$ ,  \textbf{(c)} $s_3 = (-3, -2)$ and \textbf{(d)} $s_4 = (3, -2).$ }
    \label{fig1}
\end{figure}

\begin{figure}
    \centering
    \includegraphics[width=0.5\textwidth]{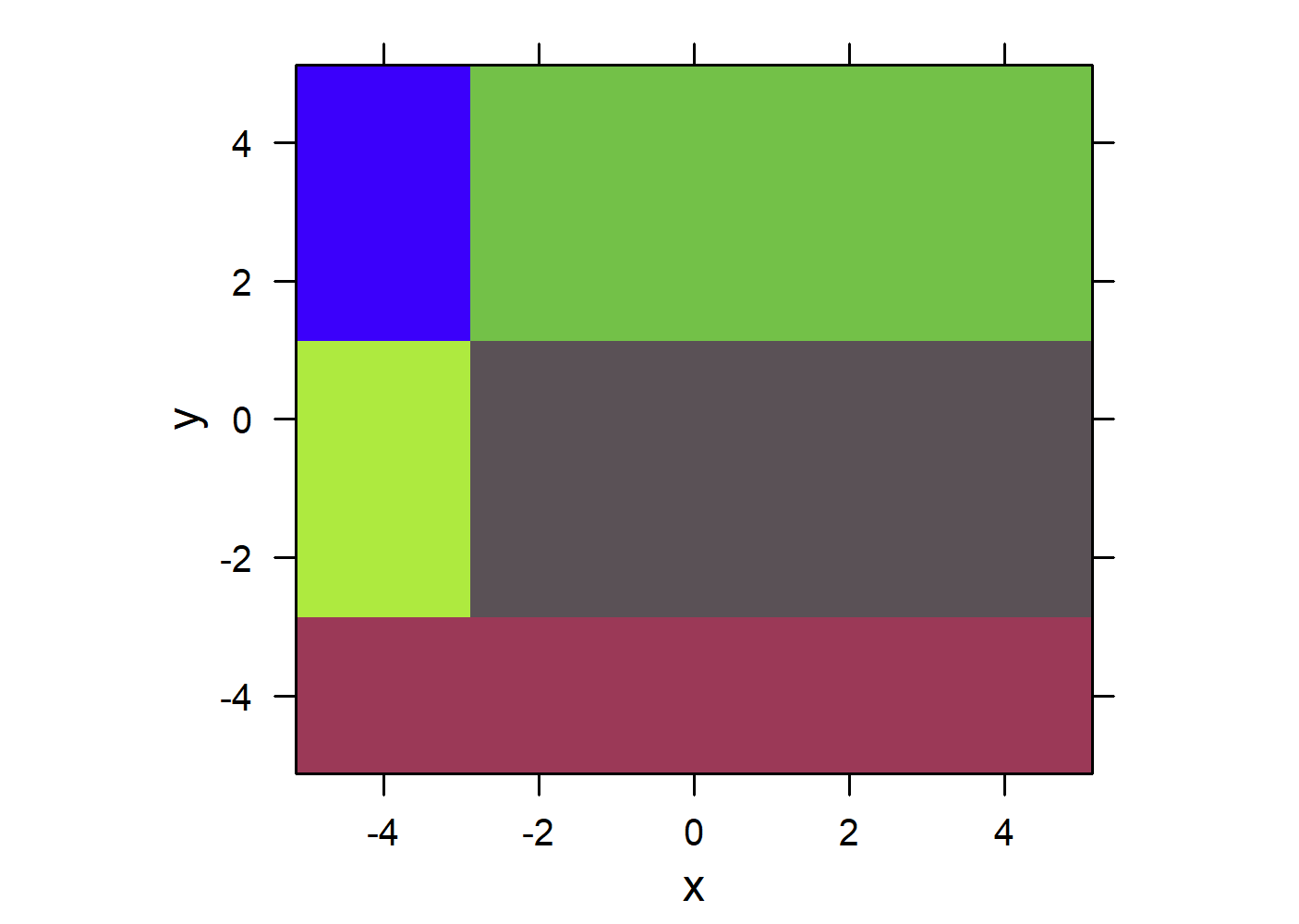}
    \caption{The results of the EDC clustering algorithm, applied to the true extremal coefficients of the non-stationary max-stable process from Fig.~\ref{fig1} using a number of clusters of five.}
    \label{fig2}
\end{figure}

As a simple example of a non-stationary process we use a Huser-Genton process on the set $\calS =[-5,5]\times[-5,5]$ with matrix parameters $a_s = 2$ constant, $b_s = (x+5)/2$, $s = (x,y) \in \calS$ and $\gamma_s = 0$ constant.  The global model parameters are $\nu = 5$ and $\alpha = 1$. This  process is obviously stationary on the sets $\{x\} \times [-5,5]$ for all $x \in [-5,5]$, and if we investigate a vertical stripe of the form $[x-\epsilon,x+\epsilon] \times [-5,5]$, $\epsilon > 0$ small, the values of $a_s$, $b_s$ and $\gamma_s$ in that region are very similar and stationarity is a reasonable approximation. The clustering based on extremal coefficients, however, does not result in clusters of such a form. In Fig.~\ref{fig1} we depict the pairwise extremal coefficients $\theta_{s,t}$, $t \in \calS$ for four selected values of $s$: $s_1 = (-3, 2)$, $s_2 = ( 3, 2)$, $s_3 = (-3, -2)$ and $s_4 = (3, -2)$. It can be observed also from Fig.~\ref{fig1} that the dependence structures for the points with the same value of $x$ are identical and that it would therefore be reasonable to group them into the same cluster. However, the extremal coefficient $\theta_{s_1,s_2}$ is close to $2$, so the points $s_1$ and $s_2$ will likely not be grouped into the same cluster by the EDC clustering. The same holds for the points $s_3$ and $s_4$. Instead, pairs of points with a low extremal coefficient, like for example $(2, 2)$ and $(3,2)$ will be grouped together even though the dependency structures around these points differ. Indeed, if we apply the EDC clustering to the true values of the extremal coefficients, we obtain the clusters shown in Fig.~\ref{fig2}, confirming the theoretical considerations we just made.


\subsection{Clustering based on local estimates}
An approach that is expected to be more suitable to finding spatially stationary regions is based on the direct comparison of the structures of extremal dependence around two points. We assume that for each point $s \in \calS$, the extremal coefficients form a locally elliptic structure, that is, for all points $t \in \calS$ in the vicinity of $s$ the extremal coefficient $\theta_{s,t}$ can be approximated by $R_\alpha(\| A_s (s-t)\|)$ for some transformation matrix $A_s$ and some $\alpha \in (0,2]$ (fixed on the whole process), with $R_\alpha(x) = \exp(-x^\alpha)$ as in Section \ref{sec:apple}. This is obviously true for stationary extremal-t processes and \cite{husergenton} show that it is also true for the non-stationary processes they designed, so we do not consider this to be a too severe restriction. Fixing a small $\epsilon > 0$ and choosing values for the global parameters $\alpha$ and $\nu$, we obtain for each of the locations $x \in \calX$ at which data are given an estimate for $A_x$ by using composite maximum likelihood on the sum of the pairwise log-likelihoods for all pairs $(x,u)$, $u \in U(x) := \{ u \in \calX \big| \, \| x-u\|<\epsilon\}$. Depending on the spatial structure of $\calX$, some isolated points may have to be excluded  beforehand to ensure that $U(x)$ is always of a sufficient size. \\

In order to identify structures in the estimates better and to reduce the influence of outliers, we apply spatial smoothing to the estimated parameters $a,b$, and $\gamma$ at every point. In our application, to smooth the values at $x\in \calX$ we used local averages on a vicinity of $x$. Other spatial smoothing algorithms, for example kernel smoothing, are also possible \citep[see for example][]{wandjones}. This results in an estimated matrix $\hat A_x$ for each location $x \in \calX$. If we group points with similar values of $\hat A_x$ into one cluster, it is reasonable to assume that within this cluster spatial stationarity holds approximately. \\

To determine which of the matrices are considered 'similar', we compare the shapes of the elliptical level sets corresponding to them using the Jaccard index. Let $L_s := \{ h \in \R^2 \big{|} \, ||\hat A_s h|| \geq 0.5 \}$. We define a dissimilarity measure as
\begin{align}
D_2(x_1,x_2) = 1- \frac {|L_{x_1} \cap L_{x_2}|} {|L_{x_1} \cup L_{x_2}|}
\end{align}
with $|\cdot|$ denoting the area. The value of $0.5$ in the definition of $L_x$ is arbitrary, any other value in $(0,1)$ would yield the same result. We will use the term Local Estimates Clustering (LEC) for this clustering from now on. \\

\subsection{Comparison of clusterings}
\label{subsec:gof}
To compare the two algorithms, we investigate which of the resulting clusterings is better suited to fitting stationary max-stable processes to it. Assume that we have fitted both clustering algorithms to some data and that we have fitted a stationary max-stable process to each of the resulting clusters. Denote the clusters of the EDC algorithm by $\{\sigma_1,\dots,\sigma_p\}$ and the clusters of the LEC algorithm by $\{\tau_1,\dots,\tau_q\}$. Each of these elements is a subset of $\calX$, the different clusters in one clustering are disjoint and their union is the whole set $\calX$. Define $\mathcal{M} = \{\sigma_1,\dots,\sigma_p,\tau_1,\dots,\tau_q\}$ as the set of all clusters. Remember that for each cluster $M \in \mathcal{M}$, the estimated parameters of the corresponding max-stable process $\hat\psi_M$ have been calculated by maximizing the composite likelihood $L_M$ from Eq.~\eqref{eq:llh_m}. \\

Note that for either clustering algorithm, max-stable processes are fitted only to the data within the same cluster, so if two points fall into two different clusters, a statistical model for their dependency is not provided. For this reason, it is not possible to compute or compare the composite likelihoods on the whole set $\calX$ from Eq.~\eqref{eq:composite_llh}, which would otherwise be the standard approach for evaluating the goodness of fit. The likelihoods of the fitted processes of the different clusters are not comparable to each other because each one of them is based on different underlying data. However, it is possible to calculate composite likelihoods on the intersections of clusters of the two clusterings, that is, on the sets $\upsilon_{ij} = \sigma_i \cap \tau_j$, provided they contain two or more elements. On each of these intersections a stationary max-stable process has been fitted for both algorithms, and by comparing the likelihoods the goodness of fit of the processes on this area can be compared. This leads to a measure for the goodness of fit on $\upsilon_{ij}$ for both algorithms:
\begin{align}
    \hat L_{i,j}^{EDC} = L_{\upsilon_{ij}}(\hat\psi_{\sigma_i}),\qquad\qquad
    \hat L_{i,j}^{LEC} = L_{\upsilon_{ij}}(\hat\psi_{\tau_j}).
\end{align}
Note that both models have the same number of parameters, so we can compare the likelihoods directly and do not need a penalty term as in the Akaike or Bayesian Information Criterion. \\

\section{Simulation study}
\label{sec:simulation}
In this section, we compare the two clustering algorithms by means of a simulation study. To do this, we simulate data from the Huser-Genton model we already investigated in Section \ref{sec:apple}. Remember that we use for this model as global parameters $\alpha=1$ and $\nu = 5$ and as parameters for the local dependencies $a_s = 2$, $b_s = (x+5)/2$ and $\gamma_s = 0$ for $s = (x,y) \in \calS$ (see Fig.~\ref{fig1}). We choose a horizontal and vertical resolution of the space $\calS = [-5,5] \times [-5,5]$ equal to $0.2$ and simulate data from processes with $250$ independent observations. \\

\begin{figure}
    \centering
    \includegraphics[width=\textwidth]{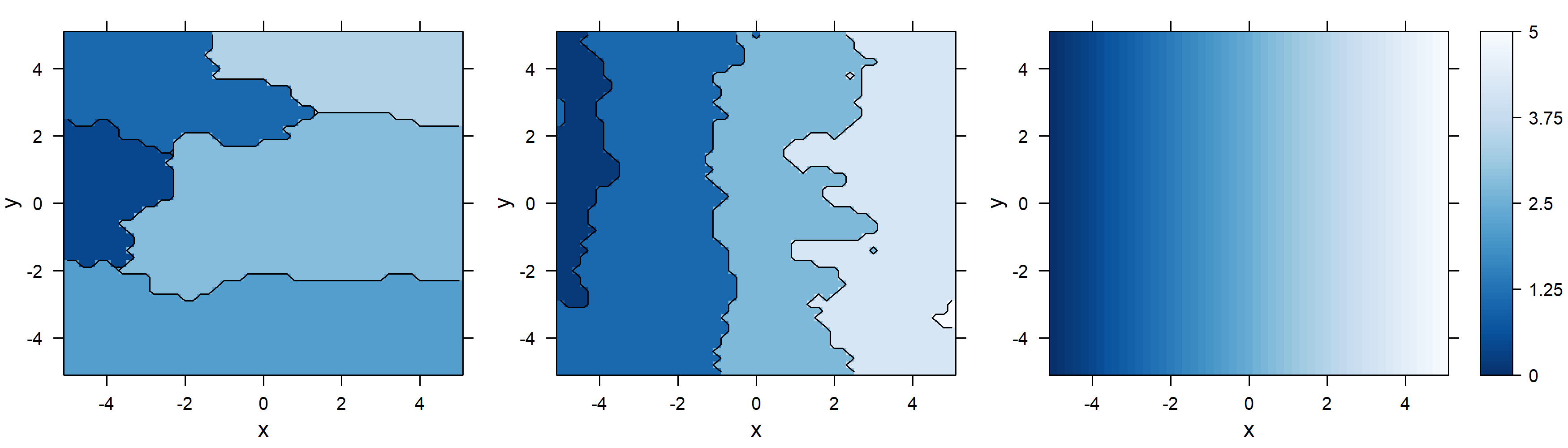}
    \caption{Results of the EDC algorithm \textbf{(a)} and the LEC algorithm \textbf{(b)} applied to simulated data of a Huser-Genton process as in Fig.~\ref{fig1} with $250$ observations. The number of clusters used is five. The colors within the clusters show the estimated value for parameter $b$ on the stationary-max-stable process that has been fitted to the data on the cluster. For reference, the true values of the parameter $b_s$ used to simulate the data are shown in \textbf{(c)}.}
    \label{fig3}
\end{figure}

For the clustering algorithms, we choose a number of clusters equal to five in both algorithms. In a first investigation, we apply the algorithms using as global parameters the true values of $\nu$ and $\alpha$. In Fig.~\ref{fig3}a and Fig.~\ref{fig3}b, the clusters produced by the two algorithms are displayed. Stationary max-stable processes are fitted to the data in the clusters, and the color inside each cluster in Fig.~\ref{fig3}a and Fig.~\ref{fig3}b depicts the value of the corresponding estimate for the parameter $b$. For reference, the true values of the parameter $b_s$ are depicted in Fig.~\ref{fig3}c. The true values of the other two parameters $a_s$ and $\gamma_s$ are constant over the whole space; their estimates are also similar for all clusters and are not shown. It can be observed that the clusters of the EDC algorithm are similar to those derived when applying the EDC algorithm using the true values (Fig.~\ref{fig2}). In particular, 
as in the theoretical case, there is considerable variation in the true values of $b_s$ within some of the clusters. A fitted stationary process cannot account for that variation. The LEC algorithm results in clusters that form vertical stripes, and on these clusters there is less variation in the true values of $b_s$. The fitted values on the clusters are therefore often closer to the true values than for the EDC algorithm (compare Fig.~\ref{fig3}b and Fig.~\ref{fig3}c). \\

\begin{figure}
    \centering
    \includegraphics[width=0.7\textwidth]{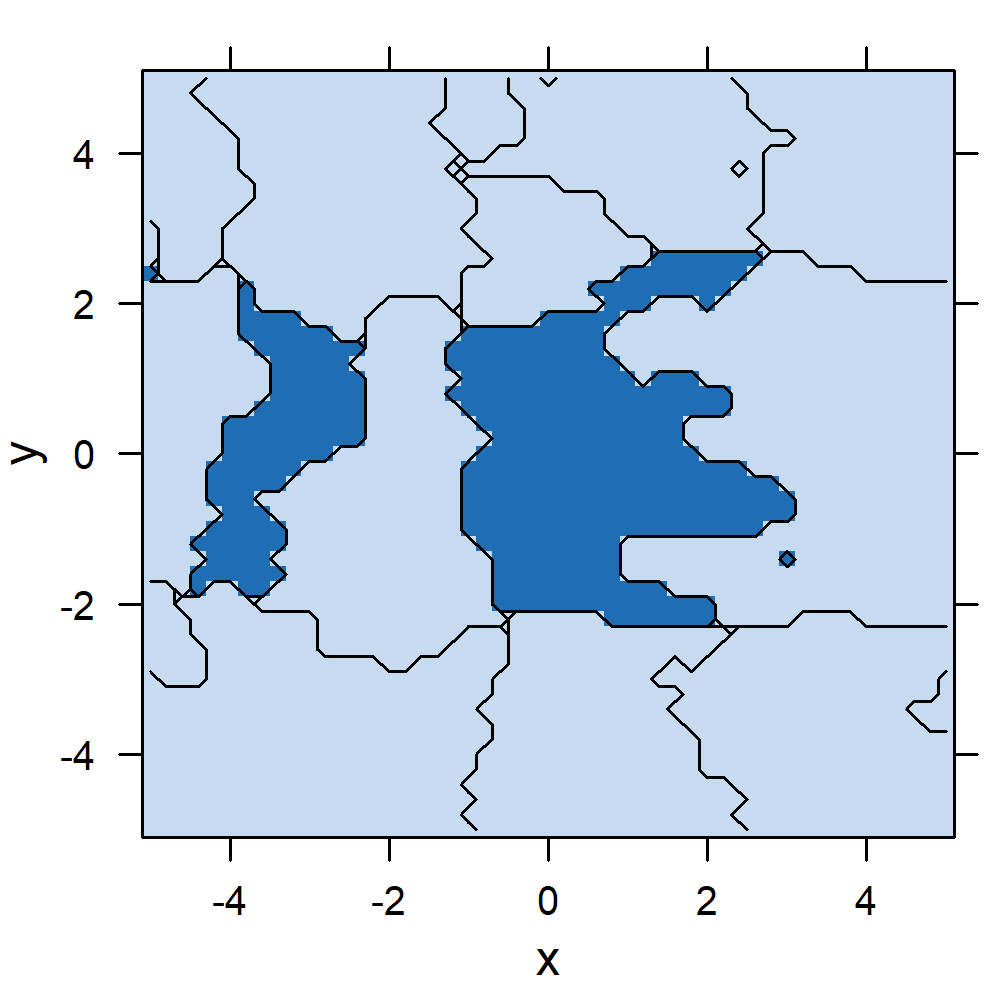}
    \caption{Comparison of the goodness of fit of the two clustering algorithms. For each intersection of the clusters of the LEC and the EDC algorithm from Fig.~\ref{fig3}, the goodness of fit of the two algorithms is compared. Dark blue color indicates that the EDC algorithm has a better goodness of fit, light blue colors indicates a better goodness of fit of the LEC algorithm.}
    \label{fig4}
\end{figure}

The two clustering algorithms are compared using the method described in Section \ref{subsec:gof}. In Fig.~\ref{fig4}, we depict the intersections of the clusters of the two algorithms. The color of each region indicates which algorithm has the better goodness of fit there (darker color \textemdash \, EDC, lighter color \textemdash \, LEC). It can be observed that the LEC algorithm results in a better goodness of fit on most regions. \\

\begin{figure}
    \centering
    \includegraphics[width=0.8\textwidth]{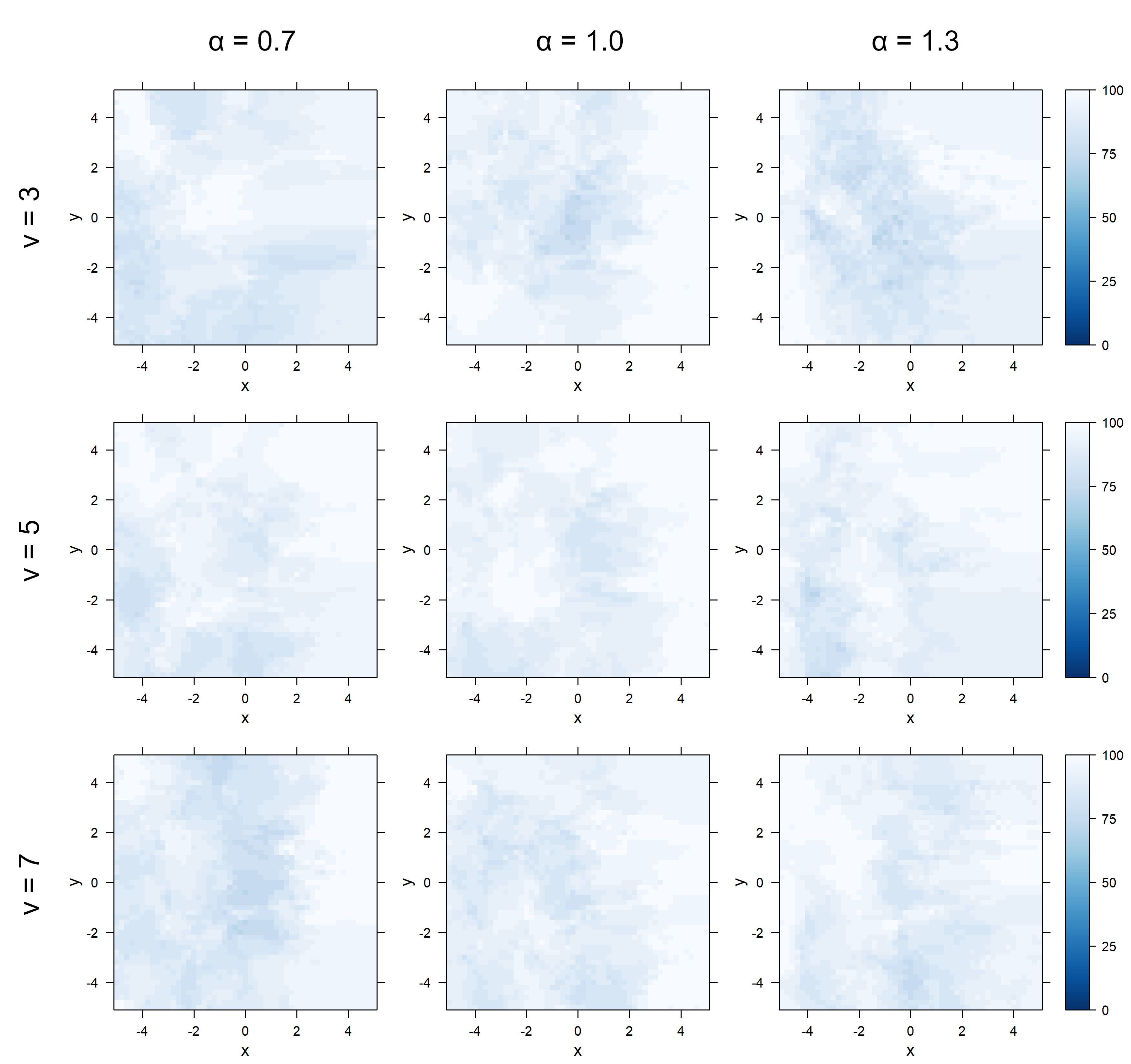}
    \caption{The analysis shown in Fig.~\ref{fig4} is repeated for $24$ independent simulations and the clustering algorithms are applied using different values for the global parameters $\nu$ and $\alpha$ (true values: $5$ and $1.0$, respectively). For each point $s \in \S$, the percentage of results for which the LEC algorithm has a better goodness of fit than EDC algorithm on the region the point is in is depicted.}
    \label{fig5}
\end{figure}

The above analyses present the results for just one simulation and using the true values for $\nu$ and $\alpha$. For a more general investigation, the simulation of a non-stationary process and its investigation is repeated another $24$ times. As the true values of $\nu$ and $\alpha$ are not known in practical applications, we conduct the analyses also for other parameter values, using for $\nu$ the values $3, 5,$ and $ 7$ and for $\alpha$ the values $0.7, 1.0, 1.3$. The resulting clusters are of course slightly different each time, but the general structures that can be identified in Fig.~\ref{fig3} stay the same (not shown). In Fig.~\ref{fig5}, for each combination of the values for $\nu$ and $\alpha$, we depict for each point $s \in \calS$ the percentage of the $24$ simulations for which the LEC algorithms has a better goodness of fit on the cluster the point is in. It can be observed that for each choice of the global parameter values and throughout the area of investigation, the LEC algorithm exhibits at each point a better goodness of fit for more than $75\%$ of the realizations. \\

\begin{figure}
    \centering
    \includegraphics[width=\textwidth]{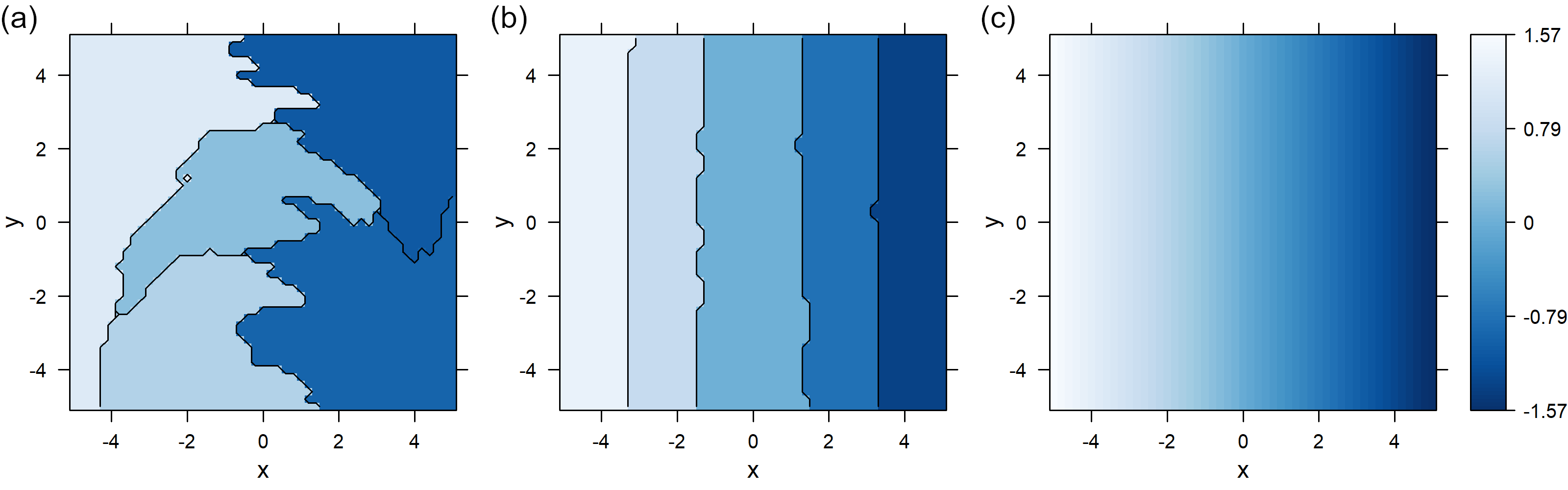}
    \caption{Results of the EDC algorithm \textbf{(a)} and the LEC algorithm \textbf{(b)} applied to simulated data of a Huser-Genton process with parameters $a_s = 1$ constant, $b_s = 3$ constant and $g_s = (-x+5) \cdot \pi/2$ with $250$ observations. The number of clusters used is five. The colors within the clusters show the estimated value for parameter $g$ on the stationary-max-stable process that has been fitted to the data on the cluster. For reference, the true values of the parameter $g_s$ are shown in \textbf{(c)}.}
    \label{fig6}
\end{figure}

As a second example for a non-stationary process we use a process with parameters $a_s = 1$ constant, $b_s = 3$ constant and $g_s = (-x+5) \cdot \pi/2$. The dependence structure around each point is locally an ellipse with the angle of the major axis rotating clockwise with increasing $x$ (vertical for $x=-5$ , horizontal for $x=0$, vertical again for $x=5$). In Fig.~\ref{fig6}a and Fig.~\ref{fig6}b we depict for one simulation the clusters produced by the two algorithms, with the colors of the clusters indicating the value of the estimate of parameter $g$ of the fitted processes. We depict the true values of parameter $g_s$ in  Fig.~\ref{fig6}c. Again, the clusters of the EDC algorithm group points with a high interdependency together, so their shape follows the rotation of the ellipses (this is visible especially well in the cluster in the middle of Fig.~\ref{fig6}a and the two clusters above it), while the LEC algorithm results in clusters that form vertical stripes and reconstruct the spatial structure in the parameter values $g_s$, $s \in \calS$ better. The analysis of the goodness of fit yields similar results as for the first algorithm, with the LEC algorithm being preferred in at least $75\%$ of the cases for every point. \\

\begin{figure}
    \centering
    \includegraphics[width=\textwidth]{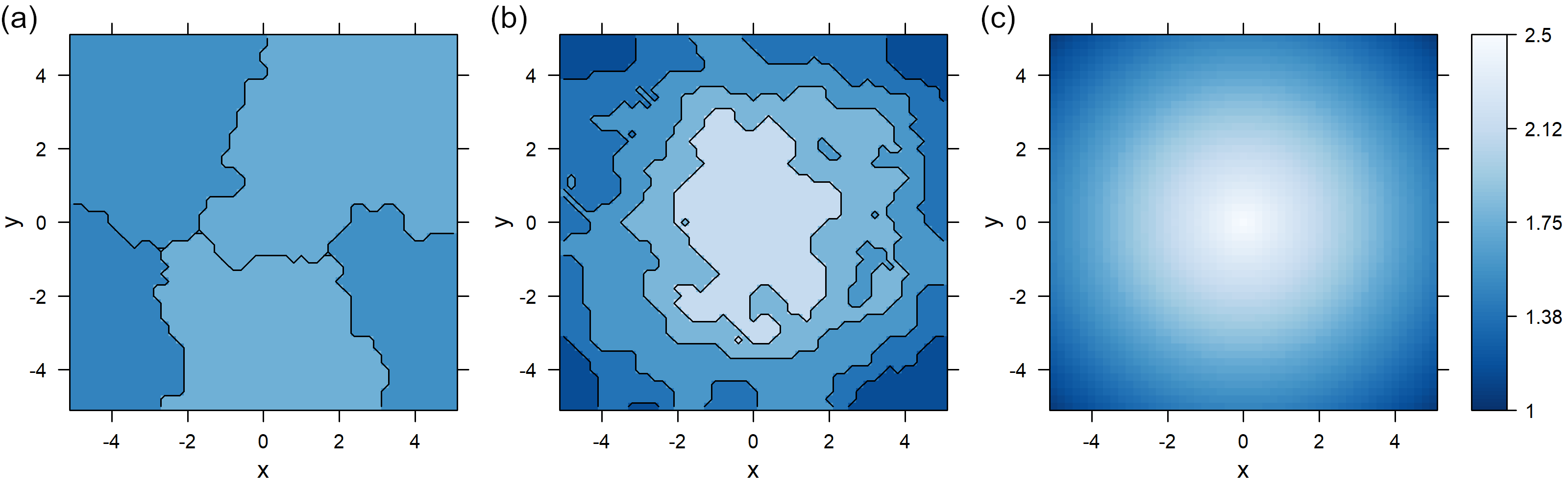}
    \caption{Results of the EDC algorithm \textbf{(a)} and the LEC algorithm \textbf{(b)} applied to simulated data of a Huser-Genton process with parameters $a_s = (7.5 - \|s\|)/2 + 1$, $b_s = 0$, $g_s = 0$ with $250$ observations. The number of clusters used is five. The colors within the clusters show the estimated value for parameter $a$ on the stationary-max-stable process that has been fitted to the data on the cluster. For reference, the true values of the parameter $a_s$ are shown in \textbf{(c)}.}
    \label{fig7}
\end{figure}

As a third example, we use values of $a_s = (7.5 - \|s\|)/2 + 1$, $b_s = 0$, $g_s = 0$. This time, the true spatial structure is a bit different, it does not feature vertical stripes, but instead a circular structure with the values of $g_s$ depending on the distance of $s$ to the center. This circular structure is visible in the clusters of the LEC algorithm (Fig.~\ref{fig7}b). It is not reproduced by the EDC algorithm, which results in clusters that are quite uninformative this time, as the fitted values for $a$ are very similar for each cluster (Fig.~\ref{fig7}a). The results for the goodness of fit are similar to those for the two examples before.

\section{Conclusions and Discussion}
\label{sec:discussion}
We have discussed a clustering algorithm by \cite{saunders} that is used in multivariate extreme value theory to group an area of investigation into smaller regions. The obtained regions are then used to fit parametric stationary max-stable processes to the data. This allows the application of such processes also to data for which stationarity on the whole area cannot be assumed. While clustering is in general a sensible and valid approach to the problem, the dissimilarity measure used previously is not necessarily suitable to find regions in which stationarity can be assumed, as we have discussed for one concrete example of a non-stationary max-stable process. We propose a different dissimilarity measure based on local estimates and demonstrate in a simulation study that for three different examples of non-stationary data we used, it indeed reconstructs the dependency structures of the data more accurately and that the processes that were fitted to the data on the clusters mostly have a better goodness of fit. For more general results regarding the performance of the two methods, a more formal and mathematically more strict analysis is required.\\

The clusters derived using the method by \cite{saunders} do have a useful and meaningful interpretation in another context: They show regions in which there is a high extremal dependency between pairs of points. Therefore, a large cluster in a certain area indicates a tendency for more large-scale extreme events in that region. Finding such clusters is of relevance for example in the context of insurances or risk management. \\

The clustering approach in this work and the subsequent fitting of max-stable processes requires choosing two global parameters, and so far, we have not found a systematic method to do so. For exactly two different choices of the values, the clusters could be calculated and the goodness of fit of the processes could be compared using the measure from Section~\ref{subsec:gof}. Unfortunately, this measure does not allow for the comparison of more than two clusterings at the same time. Carrying out a lot of pairwise comparisons is time-consuming and does not seem to be a very convenient approach. Besides, there is no guarantee that the results of these pairwise comparisons do not contradict each other. A more sophisticated approach is definitely desirable here. \\

A more general limitation with the approach of using cluster algorithms and then fitting regional max-stable models is that while those models can be used to describe the data within one cluster and also give meaningful information about how dependence structures vary spatially, they do not enable us to model directly the dependency between two points that are in different clusters. In this regard, the clustering approach is inferior to non-stationary max-stable processes like the ones presented in \cite{husergenton} (which are on the other hand more difficult to apply and require the availability of suitable covariates). It is an interesting future research direction to use the max-stable processes that have been fitted on the clusters and try to find a way to combine and extend them to a parsimonious process covering the whole area of investigation. \\

Another possible application of the clustering algorithms is the spatio-temporal investigation of extremes. In order to model changes in climate extremes over time at one specific location, a GEV distribution with time-dependent parameters can be used \citep{article2}. After fitting time-dependent models for data at different locations, the data can then be transformed to have time-stationary unit Fréchet distributions. By applying clustering algorithms to the transformed data at different time windows, it is possible to combine the temporal investigation of changes in extremes with an investigation of changes in the spatial dependence structure.

\section*{Statements and Declarations}
\begin{itemize}
\item \textbf{Funding:}
Justus Contzen is funded through the Helmholtz School for Marine Data Science (https://www.mardata.de; grant no. HIDSS-0005). Gerrit Lohmann receives funding through "Ocean and Cryosphere under climate change" in the Program "Changing Earth – Sustaining our Future" of the Helmholtz Society (https://www.helmholtz.de/en/about-us/structure-and-governance/program-oriented-funding/) and through PalMod by the Bundesministerium für Bildung und Forschung (https://www.palmod.de/; grant no. 01LP1917A). The funders had no role in study design, data collection and analysis, decision to publish, or preparation of the manuscript.
\item \textbf{Conflict of interest:}
The authors declare that they have no conflict of interest.
\item \textbf{Authors' contributions:}
All authors contributed to the study conception and design. The computer simulations were performed by Justus Contzen, who also wrote the first draft of the manuscript. Gerrit Lohmann and Thorsten Dickhaus supervised the study and commented on previous versions of the manuscript. All authors contributed to reviewing and editing the manuscript. The final manuscript was read and approved by all authors.
\end{itemize}

\bibliography{sn-bibliography}

%% BioMed_Central_Bib_Style_v1.01

\begin{thebibliography}{18}
% BibTex style file: bmc-mathphys.bst (version 2.1), 2014-07-24
\ifx \bisbn   \undefined \def \bisbn  #1{ISBN #1}\fi
\ifx \binits  \undefined \def \binits#1{#1}\fi
\ifx \bauthor  \undefined \def \bauthor#1{#1}\fi
\ifx \batitle  \undefined \def \batitle#1{#1}\fi
\ifx \bjtitle  \undefined \def \bjtitle#1{#1}\fi
\ifx \bvolume  \undefined \def \bvolume#1{\textbf{#1}}\fi
\ifx \byear  \undefined \def \byear#1{#1}\fi
\ifx \bissue  \undefined \def \bissue#1{#1}\fi
\ifx \bfpage  \undefined \def \bfpage#1{#1}\fi
\ifx \blpage  \undefined \def \blpage #1{#1}\fi
\ifx \burl  \undefined \def \burl#1{\textsf{#1}}\fi
\ifx \doiurl  \undefined \def \doiurl#1{\url{https://doi.org/#1}}\fi
\ifx \betal  \undefined \def \betal{\textit{et al.}}\fi
\ifx \binstitute  \undefined \def \binstitute#1{#1}\fi
\ifx \binstitutionaled  \undefined \def \binstitutionaled#1{#1}\fi
\ifx \bctitle  \undefined \def \bctitle#1{#1}\fi
\ifx \beditor  \undefined \def \beditor#1{#1}\fi
\ifx \bpublisher  \undefined \def \bpublisher#1{#1}\fi
\ifx \bbtitle  \undefined \def \bbtitle#1{#1}\fi
\ifx \bedition  \undefined \def \bedition#1{#1}\fi
\ifx \bseriesno  \undefined \def \bseriesno#1{#1}\fi
\ifx \blocation  \undefined \def \blocation#1{#1}\fi
\ifx \bsertitle  \undefined \def \bsertitle#1{#1}\fi
\ifx \bsnm \undefined \def \bsnm#1{#1}\fi
\ifx \bsuffix \undefined \def \bsuffix#1{#1}\fi
\ifx \bparticle \undefined \def \bparticle#1{#1}\fi
\ifx \barticle \undefined \def \barticle#1{#1}\fi
\bibcommenthead
\ifx \bconfdate \undefined \def \bconfdate #1{#1}\fi
\ifx \botherref \undefined \def \botherref #1{#1}\fi
\ifx \url \undefined \def \url#1{\textsf{#1}}\fi
\ifx \bchapter \undefined \def \bchapter#1{#1}\fi
\ifx \bbook \undefined \def \bbook#1{#1}\fi
\ifx \bcomment \undefined \def \bcomment#1{#1}\fi
\ifx \oauthor \undefined \def \oauthor#1{#1}\fi
\ifx \citeauthoryear \undefined \def \citeauthoryear#1{#1}\fi
\ifx \endbibitem  \undefined \def \endbibitem {}\fi
\ifx \bconflocation  \undefined \def \bconflocation#1{#1}\fi
\ifx \arxivurl  \undefined \def \arxivurl#1{\textsf{#1}}\fi
\csname PreBibitemsHook\endcsname

%%% 1
\bibitem[\protect\citeauthoryear{Schlather}{2002}]{schlather}
\begin{barticle}
\bauthor{\bsnm{Schlather}, \binits{M.}}:
\batitle{Models for stationary max-stable random fields}.
\bjtitle{Extremes}
\bvolume{5},
\bfpage{33}--\blpage{44}
(\byear{2002})
\doiurl{10.1023/A:1020977924878}
\end{barticle}
\endbibitem

%%% 2
\bibitem[\protect\citeauthoryear{Brown and Resnick}{1977}]{brown-resnick}
\begin{barticle}
\bauthor{\bsnm{Brown}, \binits{B.M.}},
\bauthor{\bsnm{Resnick}, \binits{S.I.}}:
\batitle{Extreme values of independent stochastic processes}.
\bjtitle{J. of Applied Probability}
\bvolume{14}(\bissue{4}),
\bfpage{732}--\blpage{739}
(\byear{1977})
\doiurl{10.2307/3213346}
\end{barticle}
\endbibitem

%%% 3
\bibitem[\protect\citeauthoryear{Kabluchko et~al.}{2009}]{kabluchko}
\begin{barticle}
\bauthor{\bsnm{Kabluchko}, \binits{Z.}},
\bauthor{\bsnm{Schlather}, \binits{M.}},
\bauthor{\bsnm{{de Haan}}, \binits{L.}}:
\batitle{Stationary max-stable fields associated to negative definite
  functions}.
\bjtitle{Annals of Probability}
\bvolume{37}(\bissue{5}),
\bfpage{2042}--\blpage{2065}
(\byear{2009})
\doiurl{10.1214/09-AOP455}
\end{barticle}
\endbibitem

%%% 4
\bibitem[\protect\citeauthoryear{Opitz}{2013}]{Opitz_2013}
\begin{barticle}
\bauthor{\bsnm{Opitz}, \binits{T.}}:
\batitle{Extremal t-processes: Elliptical domain of attraction and a spectral
  representation}.
\bjtitle{J. of Multivariate Analysis}
\bvolume{122},
\bfpage{409}--\blpage{413}
(\byear{2013})
\doiurl{10.1016/j.jmva.2013.08.008}
\end{barticle}
\endbibitem

%%% 5
\bibitem[\protect\citeauthoryear{Ribatet}{2017}]{ribatet}
\begin{bchapter}
\bauthor{\bsnm{Ribatet}, \binits{M.}}:
\bctitle{Modelling spatial extremes using max-stable processes}.
In: \beditor{\bsnm{Franzke}, \binits{C.L.E.}},
\beditor{\bsnm{O'Kane}, \binits{T.J.}} (eds.)
\bbtitle{Nonlinear and Stochastic Climate Dynamics},
pp. \bfpage{369}--\blpage{391}.
\bpublisher{Cambridge University Press},
\blocation{Cambridge}
(\byear{2017}).
\doiurl{10.1017/9781316339251.014}
\end{bchapter}
\endbibitem

%%% 6
\bibitem[\protect\citeauthoryear{Huser and Genton}{2016}]{husergenton}
\begin{barticle}
\bauthor{\bsnm{Huser}, \binits{R.}},
\bauthor{\bsnm{Genton}, \binits{M.G.}}:
\batitle{Non-stationary dependence structures for spatial extremes}.
\bjtitle{J. of Agricultural, Biological, and Environmental Statistics}
\bvolume{21},
\bfpage{470}--\blpage{491}
(\byear{2016})
\doiurl{10.1007/s13253-016-0247-4}
\end{barticle}
\endbibitem

%%% 7
\bibitem[\protect\citeauthoryear{Saunders et~al.}{2021}]{saunders}
\begin{barticle}
\bauthor{\bsnm{Saunders}, \binits{K.}},
\bauthor{\bsnm{Stephenson}, \binits{A.G.}},
\bauthor{\bsnm{Karoly}, \binits{D.J.}}:
\batitle{A regionalisation approach for rainfall based on extremal dependence}.
\bjtitle{Extremes}
\bvolume{24},
\bfpage{215}--\blpage{240}
(\byear{2021})
\doiurl{10.1007/s10687-020-00395-y}
\end{barticle}
\endbibitem

%%% 8
\bibitem[\protect\citeauthoryear{Bernard et~al.}{2013}]{bernard}
\begin{barticle}
\bauthor{\bsnm{Bernard}, \binits{E.}},
\bauthor{\bsnm{Naveau}, \binits{P.}},
\bauthor{\bsnm{Vrac}, \binits{M.}},
\bauthor{\bsnm{Mestre}, \binits{O.}}:
\batitle{Clustering of maxima: Spatial dependencies among heavy rainfall in
  france}.
\bjtitle{J. of Climate}
\bvolume{26}(\bissue{20}),
\bfpage{7929}--\blpage{7937}
(\byear{2013})
\doiurl{10.1175/JCLI-D-12-00836.1}
\end{barticle}
\endbibitem

%%% 9
\bibitem[\protect\citeauthoryear{{de Haan} and
  Ferreira}{2006}]{dehaan_ferreira}
\begin{bbook}
\bauthor{\bsnm{{de Haan}}, \binits{L.}},
\bauthor{\bsnm{Ferreira}, \binits{A.}}:
\bbtitle{Extreme Value Theory: An Introduction}.
\bpublisher{Springer},
\blocation{New {Y}ork}
(\byear{2006})
\end{bbook}
\endbibitem

%%% 10
\bibitem[\protect\citeauthoryear{McNeil et~al.}{2015}]{mcneil}
\begin{bbook}
\bauthor{\bsnm{McNeil}, \binits{A.J.}},
\bauthor{\bsnm{Frey}, \binits{R.}},
\bauthor{\bsnm{Embrechts}, \binits{P.}}:
\bbtitle{Quantitative Risk Management: Concepts, Techniques and Tools. Revised
  Edition}.
\bsertitle{Economics Books}.
\bpublisher{Princeton University Press},
\blocation{Princeton}
(\byear{2015})
\end{bbook}
\endbibitem

%%% 11
\bibitem[\protect\citeauthoryear{{de Haan}}{1984}]{dehaan}
\begin{barticle}
\bauthor{\bsnm{{de Haan}}, \binits{L.}}:
\batitle{A spectral representation for max-stable processes}.
\bjtitle{Annals of Probability}
\bvolume{12}(\bissue{4}),
\bfpage{1194}--\blpage{1204}
(\byear{1984})
\doiurl{10.1214/aop/1176993148}
\end{barticle}
\endbibitem

%%% 12
\bibitem[\protect\citeauthoryear{Penrose}{1992}]{penrose}
\begin{barticle}
\bauthor{\bsnm{Penrose}, \binits{M.D.}}:
\batitle{Semi-min-stable processes}.
\bjtitle{Annals of Probability}
\bvolume{20}(\bissue{3}),
\bfpage{1450}--\blpage{1463}
(\byear{1992})
\doiurl{10.1214/aop/1176989700}
\end{barticle}
\endbibitem

%%% 13
\bibitem[\protect\citeauthoryear{Davis et~al.}{2013}]{davis}
\begin{barticle}
\bauthor{\bsnm{Davis}, \binits{R.A.}},
\bauthor{\bsnm{Klüppelberg}, \binits{C.}},
\bauthor{\bsnm{Steinkohl}, \binits{C.}}:
\batitle{Max-stable processes for modeling extremes observed in space and
  time}.
\bjtitle{J. of the Korean Statistical Society}
\bvolume{42}(\bissue{3}),
\bfpage{399}--\blpage{414}
(\byear{2013})
\doiurl{10.1016/j.jkss.2013.01.002}
\end{barticle}
\endbibitem

%%% 14
\bibitem[\protect\citeauthoryear{Ribatet et~al.}{2015}]{ribatet_dombry}
\begin{bchapter}
\bauthor{\bsnm{Ribatet}, \binits{M.}},
\bauthor{\bsnm{Dombry}, \binits{C.}},
\bauthor{\bsnm{Oesting}, \binits{M.}}:
\bctitle{Spatial extremes and max-stable processes}.
In: \beditor{\bsnm{Dey}, \binits{D.}},
\beditor{\bsnm{Yan}, \binits{J.}} (eds.)
\bbtitle{Extreme Value Modeling and Risk Analysis: Methods and Applications},
pp. \bfpage{179}--\blpage{194}.
\bpublisher{Chapman and Hall/CRC, New York}, \blocation{???}
(\byear{2015}).
\doiurl{10.1201/b19721-10}
\end{bchapter}
\endbibitem

%%% 15
\bibitem[\protect\citeauthoryear{Cooley et~al.}{2006}]{cooley}
\begin{bchapter}
\bauthor{\bsnm{Cooley}, \binits{D.}},
\bauthor{\bsnm{Naveau}, \binits{P.}},
\bauthor{\bsnm{Poncet}, \binits{P.}}:
\bctitle{Variograms for spatial max-stable random fields}.
In: \beditor{\bsnm{Bertail}, \binits{P.}},
\beditor{\bsnm{Soulier}, \binits{P.}},
\beditor{\bsnm{Doukhan}, \binits{P.}} (eds.)
\bbtitle{Dependence in Probability and Statistics}.
\bpublisher{Springer},
\blocation{New {Y}ork}
(\byear{2006})
\end{bchapter}
\endbibitem

%%% 16
\bibitem[\protect\citeauthoryear{Paciorek and Schervish}{2006}]{paciorek}
\begin{barticle}
\bauthor{\bsnm{Paciorek}, \binits{C.J.}},
\bauthor{\bsnm{Schervish}, \binits{M.J.}}:
\batitle{Spatial modelling using a new class of nonstationary covariance
  functions}.
\bjtitle{Environmetrics}
\bvolume{17}(\bissue{5}),
\bfpage{483}--\blpage{506}
(\byear{2006})
\doiurl{10.1002/env.785}
\end{barticle}
\endbibitem

%%% 17
\bibitem[\protect\citeauthoryear{Wand and Jones}{1994}]{wandjones}
\begin{bbook}
\bauthor{\bsnm{Wand}, \binits{M.P.}},
\bauthor{\bsnm{Jones}, \binits{M.C.}}:
\bbtitle{Kernel Smoothing (1st Ed.)}.
\bpublisher{Chapman and Hall/CRC},
\blocation{New York}
(\byear{1994}).
\doiurl{10.1201/b14876}
\end{bbook}
\endbibitem

%%% 18
\bibitem[\protect\citeauthoryear{Contzen et~al.}{2023}]{article2}
\begin{barticle}
\bauthor{\bsnm{Contzen}, \binits{J.}},
\bauthor{\bsnm{Dickhaus}, \binits{T.}},
\bauthor{\bsnm{Lohmann}, \binits{G.}}:
\batitle{Long-term temporal evolution of extreme temperature in a warming
  earth}.
\bjtitle{PLOS ONE}
\bvolume{18}(\bissue{2}),
\bfpage{1}--\blpage{30}
(\byear{2023})
\doiurl{10.1371/journal.pone.0280503}
\end{barticle}
\endbibitem

\end{thebibliography}

\end{document}